\documentclass[aps,eqsecnum,preprint,floats,epsf,epsfig,nofootinbib,letter]{revtex4}
\usepackage{epsfig}

\begin{document}
\def\be{\begin{eqnarray}}
\def\en{\end{eqnarray}}
\def\non{\nonumber}
\def\ov{\overline}
\def\la{\langle}
\def\ra{\rangle}
\def\B{{\cal B}}
\def\pr{{\sl Phys. Rev.}~}
\def\prl{{\sl Phys. Rev. Lett.}~}
\def\pl{{\sl Phys. Lett.}~}
\def\np{{\sl Nucl. Phys.}~}
\def\zp{{\sl Z. Phys.}~}
\def\lsim{ {\ \lower-1.2pt\vbox{\hbox{\rlap{$<$}\lower5pt\vbox{\hbox{$\sim$}
}}}\ } }
\def\gsim{ {\ \lower-1.2pt\vbox{\hbox{\rlap{$>$}\lower5pt\vbox{\hbox{$\sim$}
}}}\ } }

\font\el=cmbx10 scaled \magstep2{\obeylines\hfill September, 2015}

\vskip 1.5 cm

\centerline{\large\bf Strong Decays of Charmed Baryons in Heavy Hadron }
\centerline{\large\bf Chiral Perturbation Theory: An Update}

\vskip 1.5 cm

\bigskip
\bigskip
\centerline{\bf Hai-Yang Cheng$^1$, Chun-Khiang Chua$^2$}
\medskip
\centerline{Institute of Physics, Academia Sinica}
\centerline{Taipei, Taiwan 115, Republic of China}
\medskip
\medskip
\centerline{$^2$ Department of Physics and Center for High Energy Physics}
\centerline{Chung Yuan Christian University}
\centerline{Chung-Li, Taiwan 320, Republic of China}

\bigskip
\bigskip
\centerline{\bf Abstract}
\bigskip
\small
We first give a brief overview of the charmed baryon spectroscopy and discuss their possible structure and spin-parity assignments in the quark model.
With the new Belle measurement of the widths of $\Sigma_c(2455)$ and $\Sigma_c(2520)$ and the recent CDF measurement of the strong decays of $\Lambda_c(2595)$ and $\Lambda_c(2625)$, we give updated coupling constants in heavy hadron chiral perturbation theory. We find $g_2=0.565^{+0.011}_{-0.024}$ for $P$-wave transitions between $s$-wave and $s$-wave baryons, and $h_2$, one of the couplings responsible for $S$-wave transitions between $s$-wave and $p$-wave baryons, is extracted from $\Lambda_c(2595)^+\to\Lambda_c^+\pi\pi$ to be $0.63\pm0.07$. It is substantially enhanced compared to the old value of order 0.437. With the help from the quark model, two of the couplings $h_{10}$ and $h_{11}$ responsible for $D$-wave transitions between $s$-wave and $p$-wave baryons are determined from $\Sigma_c(2880)$ decays.  There is a tension for the coupling $h_2$ as its value extracted from $\Lambda_c(2595)^+\to \Lambda_c^+\pi\pi$ will imply $\Xi_c(2790)^0\to\Xi'_c\pi$ and $\Xi_c(2815)^+\to\Xi_c^*\pi$ rates slightly above the current limits. It is conceivable  that SU(3) flavor symmetry breaking can help account for the discrepancy.

\pagebreak


\section{Introduction}
Many new excited charmed baryon states have been
discovered by BaBar, Belle, CLEO and LHCb in the past decade. A very rich source of charmed baryons comes both from $B$
decays and from the $e^+e^-\to c\bar c$ continuum.
Experimentally and theoretically, it is important to identify the
quantum numbers of these new states and understand their
properties. Since the pseudoscalar mesons involved in the strong
decays of charmed baryons are soft, the charmed baryon system
offers an excellent ground for testing the ideas and predictions
of heavy quark symmetry of the heavy quark and chiral symmetry of
the light quarks. The strong decays of charmed baryons are most conveniently
described by the heavy hadron chiral perturbation theory (HHChPT) in which heavy quark symmetry and chiral symmetry are incorporated
\cite{Yan,Wise}. Heavy baryon chiral Lagrangians  were first constructed in \cite{Yan} for strong decays of $s$-wave charmed baryons and in \cite{Cho,Pirjol} for $p$-wave ones. Previous phenomenological studies of the strong decays of $p$-wave charmed baryons based on HHChPT can be found in \cite{Cho,Pirjol,Chiladze,Falk03,CC}.

With the new Belle measurement of the $\Sigma_c(2455)$ and $\Sigma_c(2520)$ widths and the recent CDF measurement of the strong decays of $\Lambda_c(2595)$ and $\Lambda_c(2625)$, we would like to update the coupling constants appearing in heavy hadron chiral perturbation theory.
Indeed, this work is basically the update of \cite{CC}. We begin with the spectroscopy of charmed baryon states and discuss their possible spin-parity quantum numbers and inner structure in Sec. 2. Then in Sec. 3 we consider the strong decays of $s$-wave and $p$-wave baryons within the framework of HHChPT and update the relevant coupling constants. Sec. 4 comes to our conclusions.

\section{Spectroscopy}
Charmed baryon spectroscopy provides an ideal place for studying
the dynamics of the light quarks in the environment of a heavy
quark. The singly charmed baryon is composed of a charmed quark and
two light quarks, which we will often refer to as a diquark. Each
light quark is a triplet of the flavor SU(3). Since ${\bf 3}\times
{\bf 3}={\bf \bar 3}+{\bf 6}$, there are two different SU(3)
multiplets of charmed baryons: a symmetric sextet {\bf 6} and an
antisymmetric antitriplet ${\bf \bar 3}$. the $\Lambda_c^+$,
$\Xi_c^+$ and $\Xi_c^0$ form an ${\bf \bar 3}$ representation and
they all decay weakly. The $\Omega_c^0$, $\Xi'^+_c$, $\Xi'^0_c$
and $\Sigma_c^{++,+,0}$ form a {\bf 6} representation; among them,
only $\Omega_c^0$ decays weakly. We have followed the Particle
Data Group's convention \cite{PDG} to use a prime to distinguish the
$\Xi_c$ in the {\bf 6} from the one in the ${\bf \bar 3}$.

In the quark model, the orbital angular momentum of the light
diquark can be decomposed into ${\bf L}_\ell={\bf L}_\rho+{\bf
L}_\lambda$ (not $L_\ell=L_\rho+ L_\lambda$!\,), where ${\bf L}_\rho$ is the orbital angular momentum
between the two light quarks and ${\bf L}_\lambda$ the orbital
angular momentum between the diquark and the charmed quark. The
lowest-lying orbitally excited baryon states are the $p$-wave
charmed baryons. Denoting the quantum numbers $L_\rho$ and
$L_\lambda$ as the eigenvalues of ${\bf L}_\rho^2$ and ${\bf
L}_\lambda^2$, respectively, the $p$-wave heavy baryon can be
either in the $(L_\rho=0,L_\lambda=1)$ $\lambda$-state or the
$(L_\rho=1,L_\lambda=0)$ $\rho$-state. It is obvious that the
orbital $\lambda$-state ($\rho$-state) is symmetric (antisymmetric)
under the interchange of two light quarks $q_1$ and $q_2$. The
total angular momentum of the diquark is ${\bf J}_\ell={\bf
S}_\ell+{\bf L}_\ell$ and the total angular momentum of the charmed
baryon is ${\bf J}={\bf S}_c+{\bf J}_\ell$. In the heavy quark
limit, the spin of the charmed quark $S_c$ and the total angular
momentum of the two light quarks $J_\ell$ are separately conserved. In the following, we shall use the notation
${\cal B}_{cJ_\ell}(J^P)$ ($\tilde{\cal B}_{cJ_\ell}(J^P)$) to denote the states symmetric (antisymmetric) in the orbital wave functions under the exchange of two light quarks. The lowest-lying orbitally excited baryon states are the $p$-wave
charmed baryons with their quantum numbers listed in Table
\ref{tab:pwave}.

The next orbitally excited states are the positive-parity
excitations with $L_\rho+L_\lambda=2$.  There are multiplets
for the first positive-parity excited charmed baryons (e.g. $\Lambda_{c2}$ and $\hat\Lambda_{c2}$) with the
symmetric orbital wave function, corresponding to
$L_\lambda=2,L_\rho=0$ and $L_\lambda=0,L_\rho=2$ (see Table \ref{tab:pp}). They are distinguished by a hat.
\footnote{In our original paper \cite{CC}, we did not explicitly distinguish between $L_\lambda=2,L_\rho=0$ and $L_\lambda=0,L_\rho=2$ orbital states.}
For the case of $L_\lambda=L_\rho=1$, the total orbital angular
momentum $L_\ell$ of the diquark is 2, 1 or 0. Since the orbital
states are antisymmetric under the interchange of two light quarks,
we shall use a tilde to denote the $L_\lambda=L_\rho=1$ states. Moreover, we shall use the notation $\tilde {\cal B}^{L_\ell}_{cJ_\ell}(J^P)$ for tilde states in the ${\bf \bar 3}$ as the quantum number $L_\ell$ is needed to distinguish different states. \footnote{In terms of the old notation in \cite{CC}, $\tilde \Xi''_{c1}({1\over 2}^+,{3\over 2}^+)$ stands for $\tilde \Xi^1_{c1}({1\over 2}^+,{3\over 2}^+)$ and $\tilde \Xi'''_{c2}({3\over 2}^+,{5\over 2}^+)$ for $\tilde \Xi^2_{c2}({3\over 2}^+,{5\over 2}^+)$, for example.}

\begin{table}[t]
\caption{The $p$-wave charmed baryons denoted by ${\cal B}_{cJ_\ell}(J^P)$ and $\tilde {\cal B}_{cJ_\ell}(J^P)$
where $J_\ell$ is the total angular momentum of
the two light quarks.  In the quark model, the orbital $\lambda$-states with $L_\lambda=1$  ($\rho$-states with $L_\rho=1$) have even (odd) orbital wave functions under the
permutation of the two light quarks. The $\rho$-states are denoted by a tilde. A prime is used to distinguish
between the sextet and antitriplet SU(3) flavor states of the
$\Xi_c$.  The
explicit quark model wave functions for $p$-wave charmed baryons
can be found in \cite{Pirjol}.} \label{tab:pwave}
\begin{center}
\begin{tabular}{|c|cccc||c|cccc|} \hline
~~~~~State~~~~~ & SU(3) & ~~$S_\ell$~~ & ~~$L_\ell(L_\rho,L_\lambda)$~~&
~~$J_\ell^{P_\ell}$~~ & ~~~~~State~~~~~ & SU(3) & ~~$S_\ell$~~ &
~~$L_\ell(L_\rho,L_\lambda)$~~&
~~$J_\ell^{P_\ell}$~ \\
 \hline
 $\Lambda_{c1}({1\over 2}^{-},{3\over 2}^{-})$ & ${\bf \bar 3}$ & 0 & 1\,(0,1) &
 $1^-$ &
 $\Sigma_{c0}({1\over 2}^{-})$ & ${\bf 6}$ & 1 & 1\,(0,1) & $0^-$
 \\
 $\tilde\Lambda_{c0}({1\over 2}^{-})$ & ${\bf \bar 3}$ & 1 & 1\,(1,0) & $0^-$ &
 $\Sigma_{c1}({1\over 2}^{-},{3\over 2}^{-})$ & ${\bf 6}$ & 1 & 1\,(0,1) & $1^-$
 \\
 $\tilde\Lambda_{c1}({1\over 2}^{-},{3\over 2}^{-})$ & ${\bf \bar 3}$ & 1 & 1\,(1,0) & $1^-$
 &
 $\Sigma_{c2}({3\over 2}^{-},{5\over 2}^{-})$ & ${\bf 6}$ & 1 & 1\,(0,1) & $2^-$
 \\
 $\tilde\Lambda_{c2}({3\over 2}^{-},{5\over 2}^{-})$ & ${\bf \bar 3}$ & 1 & 1\,(1,0) & $2^-$
 &
 $\tilde \Sigma_{c1}({1\over 2}^{-},{3\over 2}^{-})$ & ${\bf 6}$ & 0 & 1\,(1,0) & $1^-$
 \\
 \hline
 $\Xi_{c1}({1\over 2}^{-},{3\over 2}^{-})$ & ${\bf \bar 3}$ & 0 & 1\,(0,1) & $1^-$  &
 $\Xi'_{c0}({1\over 2}^{-})$ & ${\bf 6}$ & 1 & 1\,(0,1) & $0^-$ \\
 $\tilde\Xi_{c0}({1\over 2}^{-})$ & ${\bf \bar 3}$ & 1 & 1\,(1,0) & $0^-$
 &  $\Xi'_{c1}({1\over 2}^{-},{3\over 2}^{-})$ & ${\bf 6}$ & 1 & 1\,(0,1) &
 $1^-$\\
 $\tilde\Xi_{c1}({1\over 2}^{-},{3\over 2}^{-})$ & ${\bf \bar 3}$ & 1 & 1\,(1,0) & $1^-$
 &  $\Xi'_{c2}({3\over 2}^{-},{5\over 2}^{-})$ & ${\bf 6}$ & 1 & 1\,(0,1) &
 $2^-$\\
 $\tilde\Xi_{c2}({3\over 2}^{-},{5\over 2}^{-})$ & ${\bf \bar 3}$ & 1 & 1\,(1,0) & $2^-$
 &  $\tilde\Xi'_{c1}({1\over 2}^{-},{3\over 2}^{-})$ & ${\bf 6}$ & 0 & 1\,(1,0) & $1^-$ \\
 \hline
\end{tabular}
\end{center}
\end{table}

\begin{table}[t]
\caption{The first positive-parity excitations of charmed baryons denoted by
${\cal B}_{cJ_\ell}(J^P)$,  $\hat {\cal B}_{cJ_\ell}(J^P)$ and $\tilde {\cal B}^{L_\ell}_{cJ_\ell}(J^P)$. States with antisymmetric orbital wave
functions (i.e. $L_\rho=L_\lambda=1$) under the interchange of two light
quarks are denoted by a tilde. States with the symmetric orbital wave
functions $L_\rho=2$ and $L_\lambda=0$ are
denoted by a hat. A prime is used to distinguish
between the sextet and antitriplet SU(3) flavor states of the
$\Xi_c$. For convenience, we drop the superscript $L_\ell$ for tilde states in the sextet.
} \label{tab:pp}
\begin{center}
\begin{tabular}{|c|cccc||c|cccc|} \hline
~~~~~State~~~~~ & SU(3)$_F$ & ~~$S_\ell$~~ & ~~$L_\ell(L_\rho,L_\lambda)$~~&
~~$J_\ell^{P_\ell}$~~ & ~~~~~State~~~~~ & SU(3)$_F$ & ~~$S_\ell$~~
& ~~$L_\ell(L_\rho,L_\lambda)$~~&
~~$J_\ell^{P_\ell}$~ \\
 \hline
 $\Lambda_{c2}({3\over 2}^+,{5\over 2}^+)$ & ${\bf \bar 3}$ & 0 & 2\,(0,2) &
 $2^+$ & $\Sigma_{c1}({1\over 2}^+,{3\over 2}^+)$ & ${\bf 6}$ & 1 & 2\,(0,2) & $1^+$ \\
 $\hat\Lambda_{c2}({3\over 2}^+,{5\over 2}^+)$ & ${\bf \bar 3}$ & 0 & 2\,(2,0) &
 $2^+$ &  $\Sigma_{c2}({3\over 2}^+,{5\over 2}^+)$ & ${\bf 6}$ & 1 & 2\,(0,2) & $2^+$ \\
 $\tilde\Lambda_{c1}({1\over 2}^+,\frac32^+)$ & ${\bf \bar 3}$ & 1 & 0\,(1,1) & $1^+$ &
 $\Sigma_{c3}({5\over 2}^+,{7\over 2}^+)$ & ${\bf 6}$ & 1 & 2\,(0,2) & $3^+$ \\
 $\tilde\Lambda^1_{c0}({1\over 2}^+)$ & ${\bf \bar 3}$ & 1 & 1\,(1,1) & $0^+$ &
 $\hat\Sigma_{c1}({1\over 2}^+,{3\over 2}^+)$ & ${\bf 6}$ & 1 & 2\,(2,0) & $1^+$  \\
 $\tilde\Lambda^1_{c1}({1\over 2}^+,{3\over 2}^+)$ & ${\bf \bar 3}$ & 1 & 1\,(1,1) & $1^+$ &
 $\hat\Sigma_{c2}({3\over 2}^+,{5\over 2}^+)$ & ${\bf 6}$ & 1 & 2\,(2,0) & $2^+$  \\
 $\tilde\Lambda^1_{c2}({3\over 2}^+,{5\over 2}^+)$ & ${\bf \bar 3}$ & 1 & 1\,(1,1) & $2^+$ &
 $\hat\Sigma_{c3}({5\over 2}^+,{7\over 2}^+)$ & ${\bf 6}$ & 1 & 2\,(2,0) & $3^+$  \\
 $\tilde\Lambda^2_{c1}({1\over 2}^+,{3\over 2}^+)$ & ${\bf \bar 3}$ & 1 & 2\,(1,1) & $1^+$ &
 $\tilde\Sigma_{c0}({1\over 2}^+)$ & ${\bf 6}$ & 0 & 0\,(1,1) & $0^+$ \\
 $\tilde\Lambda^2_{c2}({3\over 2}^+,{5\over 2}^+)$ & ${\bf \bar 3}$ & 1 & 2\,(1,1) & $2^+$ &
  $\tilde\Sigma_{c1}({1\over 2}^+,{3\over 2}^+)$ & ${\bf 6}$ & 0 & 1\,(1,1) & $1^+$ \\
 $\tilde\Lambda^2_{c3}({5\over 2}^+,{7\over 2}^+)$ & ${\bf \bar 3}$ & 1 & 2\,(1,1) & $3^+$ &
 $\tilde\Sigma_{c2}({3\over 2}^+,{5\over 2}^+)$ & ${\bf 6}$ & 0 & 2\,(1,1) & $2^+$ \\
 \hline
 $\Xi_{c2}({3\over 2}^+,{5\over 2}^+)$ & ${\bf \bar 3}$ & 0 & 2\,(0,2) &
 $2^+$ & $\Xi'_{c1}({1\over 2}^+,{3\over 2}^+)$ & ${\bf 6}$ & 1 & 2\,(0,2) & $1^+$ \\
 $\hat\Xi_{c2}({3\over 2}^+,{5\over 2}^+)$ & ${\bf \bar 3}$ & 0 & 2\,(2,0) &
 $2^+$ & $\Xi'_{c2}({3\over 2}^+,{5\over 2}^+)$ & ${\bf 6}$ & 1 & 2\,(0,2) & $2^+$ \\
 $\tilde\Xi_{c1}({1\over 2}^+,\frac32^+)$ & ${\bf \bar 3}$ & 1 & 0\,(1,1) & $1^+$ &
 $\Xi'_{c3}({5\over 2}^+,{7\over 2}^+)$ & ${\bf 6}$ & 1 & 2\,(0,2) & $3^+$ \\
 $\tilde\Xi^1_{c0}({1\over 2}^+)$ & ${\bf \bar 3}$ & 1 & 1\,(1,1) & $0^+$ &
 $\hat\Xi'_{c1}({1\over 2}^+,{3\over 2}^+)$ & ${\bf 6}$ & 1 & 2\,(2,0) & $1^+$  \\
 $\tilde\Xi^1_{c1}({1\over 2}^+,{3\over 2}^+)$ & ${\bf \bar 3}$ & 1 & 1\,(1,1) & $1^+$ &
 $\hat\Xi'_{c2}({3\over 2}^+,{5\over 2}^+)$ & ${\bf 6}$ & 1 & 2\,(2,0) & $2^+$ \\
 $\tilde\Xi^1_{c2}({3\over 2}^+,{5\over 2}^+)$ & ${\bf \bar 3}$ & 1 & 1\,(1,1) & $2^+$ &
 $\hat\Xi'_{c3}({5\over 2}^+,{7\over 2}^+)$ & ${\bf 6}$ & 1 & 2\,(2,0) & $3^+$ \\
 $\tilde\Xi^2_{c1}({1\over 2}^+,{3\over 2}^+)$ & ${\bf \bar 3}$ & 1 & 2\,(1,1) & $1^+$ &
 $\tilde\Xi'_{c0}({1\over 2}^+)$ & ${\bf 6}$ & 0 & 0\,(1,1) & $0^+$\\
 $\tilde\Xi^2_{c2}({3\over 2}^+,{5\over 2}^+)$ & ${\bf \bar 3}$ & 1 & 2\,(1,1) & $2^+$ &
  $\tilde\Xi'_{c1}({1\over 2}^+,{3\over 2}^+)$ & ${\bf 6}$ & 0 & 1\,(1,1) & $1^+$ \\
 $\tilde\Xi^2_{c3}({5\over 2}^+,{7\over 2}^+)$ & ${\bf \bar 3}$ & 1 & 2\,(1,1) & $3^+$ &
 $\tilde\Xi'_{c2}({3\over 2}^+,{5\over 2}^+)$ & ${\bf 6}$ & 0 & 2\,(1,1) & $2^+$ \\
 \hline
\end{tabular}
\end{center}
\end{table}

The observed mass spectra and decay widths of charmed baryons are
summarized in Table \ref{tab:spectrum}.
By now, the $J^P={1\over 2}^+$ and ${1\over
2}^-$ ${\bf \bar 3}$ states: ($\Lambda_c^+$, $\Xi_c^+,\Xi_c^0)$,
($\Lambda_c(2595)^+$, $\Xi_c(2790)^+,\Xi_c(2790)^0)$, ($\Lambda_c(2625)^+$, $\Xi_c(2815)^+,\Xi_c(2815)^0)$ respectively  and
$J^P={1\over 2}^+$ and ${3\over 2}^+$ ${\bf 6}$ states:
($\Omega_c,\Sigma_c,\Xi'_c$), ($\Omega_c^*,\Sigma_c^*,\Xi'^*_c$) respectively
are established. Notice that except for the parity of the lightest
$\Lambda_c^+$ and the heavier one $\Lambda_c(2880)^+$, none of the other $J^P$ quantum numbers given in
Table \ref{tab:spectrum} has been measured. One has to rely on the
quark model to determine the $J^P$ assignments.

\begin{table}[!]
\caption{Mass spectra and widths (in units of MeV) of
charmed baryons. Experimental values are taken from the Particle
Data Group \cite{PDG}. For the widths of the $\Sigma_c(2455)^{0/++}$ and $\Sigma_c(2520)^{0/++}$ baryons, we have taken into account the recent Belle measurement \cite{Belle:2014} for average. The width of $\Xi_c(2645)^+$ is taken from \cite{Belle:dc}. For $\Xi_c(3055)^0$, we quote the preliminary result from Belle \cite{Kato}.}
\label{tab:spectrum}
\begin{center}
\begin{tabular}{|c|c ccc c c c|c|} \hline \hline
~~State~~ & ~~$J^P$~~ &~$S_\ell$~ & ~$L_\ell$~ &
~$J_\ell^{P_\ell}$~ &
~~~~~~~~~Mass~~~~~~~~~ & ~~~~Width~~~~ &~Decay modes~\\
\hline
 $\Lambda_c^+$ & ${1\over 2}^+$ & 0 & 0 & $0^+$ & $2286.46\pm0.14$ & & weak  \\
 \hline
 $\Lambda_c(2595)^+$ & ${1\over 2}^-$ & 0 & 1 & $1^-$ & $2592.25\pm0.28$ &
 $2.59\pm0.56$ & $\Lambda_c\pi\pi,\Sigma_c\pi$ \\
 \hline
 $\Lambda_c(2625)^+$ & ${3\over 2}^-$ & 0 & 1 & $1^-$ & $2628.11\pm0.19$ &
 $<0.97$ & $\Lambda_c\pi\pi,\Sigma_c\pi$ \\
 \hline
 $\Lambda_c(2765)^+$ & $?^?$ & ? & ? & $?$ & $2766.6\pm2.4$ & $50$ & $\Sigma_c\pi,\Lambda_c\pi\pi$ \\
 \hline
 $\Lambda_c(2880)^+$ & ${5\over 2}^+$ & ? & ? & ? & $2881.53\pm0.35$ & $5.8\pm1.1$
 & $\Sigma_c^{(*)}\pi,\Lambda_c\pi\pi,D^0p$ \\
 \hline
 $\Lambda_c(2940)^+$ & $?^?$ & ? & ? & $?$ & $2939.3^{+1.4}_{-1.5}$ & $17^{+8}_{-6}$ &
 $\Sigma_c^{(*)}\pi,\Lambda_c\pi\pi,D^0p$ \\ \hline
 $\Sigma_c(2455)^{++}$ & ${1\over 2}^+$ & 1 & 0 & $1^+$ & $2453.98\pm0.16$ &
 $1.94^{+0.08}_{-0.16}$ & $\Lambda_c\pi$ \\
 \hline
 $\Sigma_c(2455)^{+}$ & ${1\over 2}^+$ & 1 & 0 & $1^+$ & $2452.9\pm0.4$ &
 $<4.6$ & $\Lambda_c\pi$\\
 \hline
 $\Sigma_c(2455)^{0}$ & ${1\over 2}^+$ & 1 & 0 & $1^+$ & $2453.74\pm0.16$
 & $1.87^{+0.09}_{-0.17}$ & $\Lambda_c\pi$ \\
 \hline
 $\Sigma_c(2520)^{++}$ & ${3\over 2}^+$ & 1 & 0 & $1^+$ & $2517.9\pm0.6$
 & $14.8^{+0.3}_{-0.4}$ & $\Lambda_c\pi$\\
 \hline
 $\Sigma_c(2520)^{+}$ & ${3\over 2}^+$ & 1 & 0 & $1^+$ & $2517.5\pm2.3$
 & $<17$ & $\Lambda_c\pi$ \\
 \hline
 $\Sigma_c(2520)^{0}$ & ${3\over 2}^+$ & 1 & 0 & $1^+$ & $2518.8\pm0.6$
 & $15.3^{+0.3}_{-0.4}$ & $\Lambda_c\pi$ \\
 \hline
 $\Sigma_c(2800)^{++}$ & $?^?$ & ? & ? & ? & $2801^{+4}_{-6}$ & $75^{+22}_{-17}$ &
 $\Lambda_c\pi,\Sigma_c^{(*)}\pi,\Lambda_c\pi\pi$ \\
 \hline
 $\Sigma_c(2800)^{+}$ & $?^?$ & ? & ? & ? & $2792^{+14}_{-~5}$ & $62^{+60}_{-40}$ &
 $\Lambda_c\pi,\Sigma_c^{(*)}\pi,\Lambda_c\pi\pi$ \\
 \hline
 $\Sigma_c(2800)^{0}$ & $?^?$ & ? & ? & ? & $2806^{+5}_{-7}$ & $72^{+22}_{-15}$ &
 $\Lambda_c\pi,\Sigma_c^{(*)}\pi,\Lambda_c\pi\pi$\\
 \hline
 $\Xi_c^+$ & ${1\over 2}^+$ & 0 & 0 & $0^+$ & $2467.8^{+0.4}_{-0.6}$ & & weak \\ \hline
 $\Xi_c^0$ & ${1\over 2}^+$ & 0 & 0 & $0^+$ & $2470.88^{+0.34}_{-0.80}$ & & weak \\ \hline
 $\Xi'^+_c$ & ${1\over 2}^+$ & 1 & 0 & $1^+$ & $2575.6\pm3.1$ & & $\Xi_c\gamma$ \\ \hline
 $\Xi'^0_c$ & ${1\over 2}^+$ & 1 & 0 & $1^+$ & $2577.9\pm2.9$ & & $\Xi_c\gamma$ \\ \hline
 $\Xi_c(2645)^+$ & ${3\over 2}^+$ & 1 & 0 & $1^+$ & $2645.9^{+0.5}_{-0.6}$ & $2.6\pm0.5$ & $\Xi_c\pi$ \\
 \hline
 $\Xi_c(2645)^0$ & ${3\over 2}^+$ & 1 & 0 & $1^+$ & $2645.9\pm0.9$ & $<5.5$ & $\Xi_c\pi$ \\
 \hline
 $\Xi_c(2790)^+$ & ${1\over 2}^-$ & 0 & 1 & $1^-$ & $2789.9\pm3.2$ & $<15$ & $\Xi'_c\pi$\\
 \hline
 $\Xi_c(2790)^0$ & ${1\over 2}^-$ & 0 & 1 & $1^-$ & $2791.8\pm3.3$ & $<12$ & $\Xi'_c\pi$ \\
 \hline
 $\Xi_c(2815)^+$ & ${3\over 2}^-$ & 0 & 1 & $1^-$ & $2816.6\pm0.9$ & $<3.5$ & $\Xi^*_c\pi,\Xi_c\pi\pi,\Xi_c'\pi$ \\
 \hline
 $\Xi_c(2815)^0$ & ${3\over 2}^-$ & 0 & 1 & $1^-$ & $2819.6\pm1.2$ & $<6.5$ & $\Xi^*_c\pi,\Xi_c\pi\pi,\Xi_c'\pi$ \\
 \hline
$\Xi_c(2930)^0$ & $?^?$ & ? & ? & $?$ & $2931\pm6$ & $36\pm13$
 & $\Lambda_c \ov K$ \\
 \hline
 $\Xi_c(2980)^+$ & $?^?$ & ? & ? & $?$ & $2971.4\pm3.3$ & $26\pm7$
 & $\Sigma_c \ov K,\Lambda_c \ov K\pi,\Xi_c\pi\pi$  \\
 \hline
 $\Xi_c(2980)^0$ & $?^?$ & ? & ? & $?$ & $2968.0\pm2.6$ & $20\pm7$
 & ~$\Sigma_c \ov K,\Lambda_c \ov K\pi,\Xi_c\pi\pi$~ \\
 \hline
 $\Xi_c(3055)^+$ & $?^?$ & ? & ? & $?$ & $3054.2\pm1.3$ & $17\pm13$ &
 $\Sigma_c \ov K,\Lambda_c \ov K\pi,D\Lambda$  \\
 \hline
 $\Xi_c(3055)^0$ & $?^?$ & ? & ? & $?$ & $3059.7\pm0.8$ & $7.4\pm3.9$ &
 $\Sigma_c \ov K,\Lambda_c \ov K\pi,D\Lambda$  \\
 \hline
 $\Xi_c(3080)^+$ & $?^?$ & ? & ? & $?$ & $3077.0\pm0.4$ & $5.8\pm1.0$ &
 $\Sigma_c \ov K,\Lambda_c \ov K\pi,D\Lambda$  \\
 \hline
 $\Xi_c(3080)^0$ & $?^?$ & ? & ? & $?$ & $3079.9\pm1.4$ & $5.6\pm2.2$
 & $\Sigma_c \ov K,\Lambda_c \ov K\pi,D\Lambda$ \\
 \hline
$\Xi_c(3123)^+$ & $?^?$ & ? & ? & $?$ & $3122.9\pm1.3$ & $4.4\pm3.8$
 & $\Sigma_c^* \ov K,\Lambda_c \ov K\pi$ \\
 \hline
 $\Omega_c^0$ & ${1\over 2}^+$ & 1 & 0 & $1^+$ & $2695.2\pm1.7$ & & weak \\
 \hline
 $\Omega_c(2770)^0$ & ${3\over 2}^+$ & 1 & 0 & $1^+$ & $2765.9\pm2.0$ & & $\Omega_c\gamma$ \\
 \hline \hline
\end{tabular}
\end{center}
\end{table}

\vskip 0.2cm
In the following we discuss some of the excited charmed baryon
states:

\subsection{$\Lambda_c$ states}

There are seven lowest-lying $p$-wave $\Lambda_c$ states arising from
combining the charmed quark spin $S_c$ with light constituents in
$J_\ell^{P_\ell}=1^-$ state:  $\Lambda_{c1}({1\over 2}^-,{3\over
2}^-),\tilde\Lambda_{c1}({1\over 2}^-,{3\over
2}^-),\tilde\Lambda_{c2}({3\over 2}^-,{5\over 2}^-)$ and one
singlet $\tilde\Lambda_{c0}({1\over 2}^-)$ (see Table \ref{tab:pwave}).
$\Lambda_c(2595)^+$ and $\Lambda_c(2625)^+$ form a doublet
$\Lambda_{c1}({1\over 2}^-,{3\over 2}^-)$. The allowed strong decays are
$\Lambda_{c1}(1/2^-)\to[\Sigma_c\pi]_S,~[\Sigma_c^*\pi]_D$ and $\Lambda_{c1}(3/2^-)\to [\Sigma_c\pi]_D,~[\Sigma_c^*\pi]_{S,D},~[\Lambda_c\pi\pi]_P$.

$\Lambda_c(2765)^+$ is a broad state ($\Gamma\approx 50$ MeV)
first seen in $\Lambda_c^+\pi^+\pi^-$ by CLEO
\cite{CLEO:Lamc2880}. It appears to resonate through $\Sigma_c$
and probably also $\Sigma_c^*$. It has a nickname $\Sigma_c(2765)^+$ because whether it is a $\Lambda_c^+$ or a $\Sigma_c^+$ and whether the width might be due to overlapping states are not known. It could be a first positive-parity
excitation of $\Lambda_c$. It has also been proposed in the diquark model \cite{Ebert:2007} to be either the first radial ($2S$)
excitation of the $\Lambda_c$ with $J^P=\frac12^-$ containing the light scalar diquark or the first orbital
excitation ($1P$) of the $\Sigma_c$ with $J^P=\frac32^-$ containing the light axial-vector diquark.

The state $\Lambda_c(2880)^+$ first observed by CLEO
\cite{CLEO:Lamc2880} in $\Lambda_c^+\pi^+\pi^-$ was also seen by
BaBar in the $D^0p$ spectrum \cite{BaBar:Lamc2940}. Belle has studied the
experimental constraint on it spin-parity quantum numbers \cite{Belle:Lamc2880} and found that $J^P=\frac52^+$ is favored by the angular analysis of
$\Lambda_c(2880)^+\to\Sigma_c^{0,++}\pi^\pm$ together with the ratio of $\Sigma^*\pi/\Sigma\pi$ measured to be
\begin{eqnarray} \label{eq:R}
 R\equiv {\Gamma(\Lambda_c(2880)\to\Sigma_c^*\pi^\pm)\over
 \Gamma(\Lambda_c(2880)\to\Sigma_c\pi^\pm)}=(24.1\pm6.4^{+1.1}_{-4.5})\%.
\end{eqnarray}
In the quark
model, the candidates for the parity-even spin-${5\over 2}$ state are
$\Lambda_{c2}(\frac52^+)$, $\hat\Lambda_{c2}(\frac52^+)$,
$\tilde\Lambda^1_{c2}(\frac52^+)$,
$\tilde\Lambda^2_{c2}(\frac52^+)$ and
$\tilde\Lambda^2_{c3}(\frac52^+)$ (see Table \ref{tab:pp}), recalling that the superscript refers to the orbital angular momentum $L_\ell$ of the diquark.
Based on heavy quark symmetry alone, one cannot predict the ratio $R$ for these states except  $\tilde\Lambda^2_{c3}(\frac52^+)$.  Its decays to
$\Sigma_c^*\pi$, $\Sigma_c\pi$ and $\Lambda_c\pi$ all in $F$ waves.
It turns out that \cite{CC}
\begin{eqnarray}
 {\Gamma\left(\tilde\Lambda^2_{c3}(5/2^+)\to
[\Sigma_c^*\pi]_F\right)\over \Gamma\left(\tilde\Lambda^2_{c3}(5/2^+)\to
[\Sigma_c\pi]_F\right)}&=&{5\over
4}\,{p_\pi^7(\Lambda_c(2880)\to\Sigma_c^*\pi)\over
p_\pi^7(\Lambda_c(2880)\to\Sigma_c\pi)} \nonumber \\
&=& {5\over 4}\times 0.29=0.36\,,
\end{eqnarray}
where the factor of 5/4 follows from heavy quark symmetry.
Although this deviates from the experimental measurement
(\ref{eq:R}) by $1\sigma$, it is a robust prediction.

It is worth mentioning that the Peking group
\cite{Zhu} has studied the strong decays of charmed baryons based
on the so-called $^3P_0$ recombination model. For the
$\Lambda_c(2880)^+$, Peking group found that all the symmetric states $\Lambda_{c2}$ and $\hat\Lambda_{c2}$ are ruled out as they do not decay into  $D^0p$ according to the $^3P_0$ model. Moreover, the predicted ratio $R$ is either too large or too small compared to experiment. Therefore, it appears that the $L=2$ orbitally excited state $\tilde\Lambda^2_{c3}(\frac52^+)$ dictates the inner structure of $\Lambda_c(2880)^+$.


\subsection{$\Sigma_c$ states}
The highest isotriplet charmed baryons $\Sigma_c(2800)^{++,+,0}$
decaying to $\Lambda_c^+\pi$ were first measured by Belle
\cite{Belle:Sigc2800}. The measured widths of order 70 MeV are shown in Table \ref{tab:spectrum}. The possible quark states are $\Sigma_{c0}({1\over 2}^-)$,  $\Sigma_{c1}({1\over 2}^-,{3\over 2}^-)$, $\tilde\Sigma_{c1}({1\over 2}^-,{3\over 2}^-)$ and $\Sigma_{c2}({3\over 2}^-,{5\over 2}^-)$ (see Table \ref{tab:pwave}). Obviously, the mass analysis alone is not adequate to fix
the quantum numbers $J^P$ of $\Sigma_c(2800)$ and the study of its strong decays is necessary.
The states $\Sigma_{c1}$ and $\tilde\Sigma_{c1}$ are ruled out because their decays to $\Lambda_c^+\pi$ are excluded in the heavy quark limit. They decay mainly to the two pion system $\Lambda_c\pi\pi$ in a
$P$-wave.   Now the
$\Sigma_{c2}({3\over 2}^-,{5\over 2}^-)$ baryon decays principally into the
$\Lambda_c\pi$ system in a $D$-wave, while $\Sigma_{c0}({1\over 2}^-)$  decays into $\Lambda_c\pi$ in an $S$-wave. Since HHChPT implies a very broad $\Sigma_{c0}$ with width of order 885 MeV (see Sec.III.B below), this  $p$-wave state is also excluded. Therefore, $\Sigma_c(2800)^{++,+,0}$ are likely to
be either $\Sigma_{c2}({3\over 2}^-)$ or $\Sigma_{c2}({5\over 2}^-)$ or their mixing. In the quark-diquark model \cite{Ebert:2011}, both of them have very close masses compatible with experiment. However, if we consider the Regge trajectory in the $(J,M^2)$ plane, $\Sigma_c(2800)$ with $J^P=3/2^-$ fits nicely to the parent $\Sigma_c$ trajectory (see Fig. 2(a) of \cite{Ebert:2011}). Hence, we will advocate a $\Sigma_{c2}(3/2^-)$ quark state for $\Sigma_c(2800)$. It is worth mentioning that for light strange baryons, the first orbital excitation of the $\Sigma$ also has the quantum numbers $J^P=3/2^-$ \cite{PDG}.

\subsection{$\Xi_c$ states}
There are seven lowest-lying $p$-wave $\Xi_c$ states in the ${\bf \bar 3}$:
$\tilde\Xi_{c0}({1\over 2}^-)$, $\Xi_{c1}({1\over 2}^-$,${3\over
2}^-)$, $\tilde\Xi_{c1}({1\over 2}^-,{3\over
2}^-)$, $\tilde\Xi_{c2}({3\over 2}^-,{5\over 2}^-)$, and seven states  in the ${\bf 6}$:  $\Xi'_{c0}({1\over 2}^-)$,  $\Xi'_{c1}({1\over 2}^-,{3\over 2}^-)$, $\tilde\Xi'_{c1}({1\over 2}^-,{3\over 2}^-)$, $\Xi'_{c2}({3\over 2}^-,{5\over 2}^-)$. The states $\Xi_c(2790)$ and $\Xi_c(2815)$ form a doublet
$\Xi_{c1}({1\over 2}^-,{3\over 2}^-)$. Their strong decays are $\Xi_{c1}(1/2^-)\to [\Xi'_c\pi]_S$ and $\Xi_{c1}(3/2^-)\to [\Xi_c^*\pi]_S$ where $\Xi_c^*$ stands for $\Xi_c(2645)$.

The charmed strange baryons $\Xi_c(2980)$ and $\Xi_c(3080)$
that decay into $\Lambda_c^+K^-\pi^+$ and $\Lambda_c^+K_S^0\pi^-$ were first observed by Belle
\cite{Belle:Xic2980} and confirmed by BaBar \cite{BaBar:Xic2980}.
In the same paper, BaBar also claimed evidence of two new resonances $\Xi_c(3055)^+$ and $\Xi_c(3123)^+$. The former was confirmed by Belle, while no signature of the latter was seen \cite{Belle:dc}. The neutral $\Xi_c(3055)^0$ was observed recently by Belle in $\Lambda D^0$ decays \cite{Kato}. Another state $\Xi_c(2930)^0$  omitted from the PDG summary table has been only seen by BaBar in the $\Lambda_c^+K^-$ mass projection of $B^-\to\Lambda_c^+\bar\Lambda_c^-K^-$ \cite{BaBar:Xic2930}.

The charmed baryons $\Xi_c(2980)$, $\Xi_c(3055)$, $\Xi_c(3080)$ and $\Xi_c(3123)$ could be the first positive-parity excitations of the $\Xi_c$.
The study of the Regge phenomenology is very useful for the $J^P$ assignment of charmed baryons \cite{Ebert:2011,Guo}.
Just as the two states $\Lambda_c(2880)(5/2^+)$ and $\Lambda_c(2625)(3/2^-)$ fit nicely the parent $\Lambda_c$ Regge trajectory in the $(J,M^2)$ plane, $\Xi_c(3080)$ and $\Xi_c(2815)(3/2^-)$ fall into the parent $\Xi_c$ Regge trajectory (see Fig. 3(a) of \cite{Ebert:2011}). Hence, this suggests that $\Xi_c(3080)$ has $J^P=5/2^+$. Likewise, $\Xi_c(3055)$ with $3/2^+$ fits to the parent $\Xi_c(2790)(1/2^-)$ Regge trajectory (see Fig. 3(b) of \cite{Ebert:2011}).

Since the mass difference between the
antitriplets $\Lambda_c$ and $\Xi_c$ for
$J^P=\frac12^+,\frac12^-,\frac32^-$ is of order $180\sim 200$ MeV,
it is conceivable that $\Xi_c(2980)$ and $\Xi_c(3080)$ are the
counterparts of $\Lambda_c(2765)$ and $\Lambda_c(2880)$,
respectively, in the strange charmed baryon sector. As noted in
passing, the state $\Lambda_c(2765)^+$ could be a radial excitation $(2S)$ of $\Lambda_c^+$ and $\Lambda_c(2880)$ has
the quantum numbers $J^P={5\over 2}^+$, it is thus tempting to
assign $J^P=\frac12^+$  to $\Xi_c(2980)$ with first radial excitation and $\frac52^+$ to
$\Xi_c(3080)$. From Table \ref{tab:3and6} we see that $\Xi_c(3080)$ and $\Lambda_c(2880)$ form nicely a $J^P=5/2^+$ antitriplet as the mass difference between $\Xi_c(3080)$ and $\Lambda_c(2880)$ is consistent with that observed in other antitriplets.

In the relativistic quark-diquark model \cite{Ebert:2011},  $\Xi_c(2980)$ is a sextet $J^P={1\over 2}^+$ state. According to Table \ref{tab:pp}, possible sextet candidates are $\Xi'_{c1}({1\over 2}^+),\hat\Xi'_{c1}({1\over 2}^+),\tilde\Xi'_{c0}({1\over 2}^+)$, and $\tilde\Xi'_{c1}({1\over 2}^+)$,
recalling that a tilde is to denote states
with antisymmetric orbital wave functions (i.e. $L_\rho=L_\lambda=1$) under
the interchange of two light quarks and a hat for $L_\rho=2$ and $L_\lambda=0$ states. Strong decays of these four
states have been studied in \cite{Zhu} using the $^3P_0$
model. It turns out that $\Gamma(\tilde\Xi'_{c0}({1\over
2}^+))\approx 2.0$ MeV is too small compared to the experimental value of order 25 MeV (see Table \ref{tab:spectrum}),   while
$\hat\Xi'_{c1}({1\over 2}^+)$ yields too large $\Lambda_c^+\ov K$ and $\Xi_c\pi$ rates. In the $^3P_0$ model, the strong decay of $\Xi'_{c1}({1\over 2}^+)$ to $\Sigma_c\ov K$ is largely suppressed relative to $\Lambda_c^+\ov K$. This is not favored by experiment as the decay modes $\Lambda_c^+\ov K\pi$, $\Sigma_c\ov K$, $\Xi_c\pi\pi$ and $\Xi_c(2645)\pi$ of $\Xi_c(2980)$ have been seen, but not $\Lambda_c^+\ov K$. $\tilde\Xi'_{c1}({1\over 2}^+)$ does not decay to $\Xi_c\pi$ and $\Lambda_c\ov K$ and has a width of 28 MeV consistent with experiment.
Therefore, the favored candidate for $\Xi_c(2980)$ is
$\tilde\Xi'_{c1}({1\over 2}^+)$ which has $J_\ell=L_\ell=1$.

Just as $\Lambda_c(2880)$, $\Xi_c(3080)$ is mostly likely an antitriplet $J^P={5\over 2}^+$ state as noted in passing. The possible quark states are $\Xi_{c2}({5\over 2}^+)$, $\hat\Xi_{c2}({5\over 2}^+)$, $\tilde \Xi^1_{c2}({5\over 2}^+)$, $\tilde \Xi^2_{c2}({5\over 2}^+)$ and $\tilde \Xi^2_{c3}({5\over 2}^+)$ (see Table \ref{tab:pp}).  Since
$\Xi_c(3080)$ is above the $D\Lambda$ threshold, the two-body mode $D\Lambda$ should exist though it has not been searched for in the $D\Lambda$ spectrum. Recall that the neutral $\Xi_c(3055)^0$ was observed recently by Belle in the $D^0\Lambda$ spectrum \cite{Kato}. According to the $^3P_0$ model, the first four quark states are excluded as they do not decay into $D\Lambda$ \cite{Zhu}. The only possibility left is $\tilde \Xi^2_{c3}({5\over 2}^+)$. Although it can decay into $D\Lambda$, the identification of $\tilde \Xi^2_{c3}({5\over 2}^+)$ with $\Xi_c(3080)$ encounters two potential difficulties:
(i) its width is dominated by $\Xi_c\pi$ and $\Lambda_c^+\ov K$ modes which have not been seen experimentally, and (ii) the predicted width of order 47 MeV \cite{Zhu} is too large compared to the measured one of order 5.7 MeV, even though one may argue that the $^3P_0$ model's prediction can be easily off by a factor of $2\sim 3$ from the experimental measurement due to its inherent uncertainties \cite{Zhu}.

\begin{table}[t]
\caption{Antitriplet and sextet states of charmed baryons. The spin-parity quantum numbers of $\Xi_c(3080)$ are not yet established.
Mass differences $\Delta m_{\Xi_c\Lambda_c}\equiv m_{\Xi_c}-m_{\Lambda_c}$, $\Delta m_{\Xi'_c\Lambda_c}\equiv m_{\Xi'_c}-m_{\Lambda_c}$, $\Delta m_{\Omega_c\Xi'_c}\equiv m_{\Omega_c}-m_{\Xi'_c}$ are in units of MeV. } \label{tab:3and6}
\begin{center}
\begin{tabular}{|c| ccc |} \hline\hline
  & $\B_{cJ_\ell}(J^P)$ & States & Mass difference  \\
 \hline
 ~~${\bf \bar 3}$~~ & ~~$\B_{c0}({1\over 2}^+)$~~ &  $\Lambda_c(2287)^+$, $\Xi_c(2470)^+,\Xi_c(2470)^0$ & ~~$\Delta m_{\Xi_c\Lambda_c}=183$ ~~  \\
 & ~~$\B_{c1}({1\over 2}^-)$~~ &  $\Lambda_c(2595)^+$, $\Xi_c(2790)^+,\Xi_c(2790)^0$ & $\Delta m_{\Xi_c\Lambda_c}=198$  \\
 & ~~$\B_{c1}({3\over 2}^-)$~~ &  $\Lambda_c(2625)^+$, $\Xi_c(2815)^+,\Xi_c(2815)^0$ & $\Delta m_{\Xi_c\Lambda_c}=190$  \\
 & ~~$\tilde\B_{c3}^2({5\over 2}^+)$~~ &  $\Lambda_c(2880)^+$, $\Xi_c(3080)^+,\Xi_c(3080)^0$ & $\Delta m_{\Xi_c\Lambda_c}=196$  \\
 \hline
 ~~${\bf 6}$~~ & ~~$\B_{c1}({1\over 2}^+)$~~ &  $\Omega_c(2695)^0$, $\Xi'_c(2575)^{+,0},\Sigma_c(2455)^{++,+,0}$ & ~~~~$\Delta m_{\Xi'_c\Sigma_c}=124$, $\Delta m_{\Omega_c\Xi'_c}=119$~~  \\
 & ~~$\B_{c1}({3\over 2}^+)$~~ &  $\Omega_c(2770)^0$, $\Xi'_c(2645)^{+,0},\Sigma_c(2520)^{++,+,0}$ & ~~~~$\Delta m_{\Xi'_c\Sigma_c}=128$, $\Delta m_{\Omega_c\Xi'_c}=120$~~  \\
 \hline\hline
\end{tabular}
\end{center}
\end{table}

\subsection{$\Omega_c$ states}
Only two ground states have been observed thus far: $1/2^+$ $\Omega_c^0$ and $3/2^+$ $\Omega_c(2770)^0$. The latter was seen by BaBar in the electromagnetic decay
$\Omega_c(2770)\to\Omega_c\gamma$ \cite{BaBar:Omegacst}.

\vskip 0.5cm
Charmed baryon spectroscopy has been studied extensively in
various models. The interested readers are referred to
\cite{Chen:2015k,theorycharmspect,theorycharmspect2} for further references. It appears that  the spectroscopy is well described by the model based on the relativitsic heavy quark-light diquark model by Ebert, Faustov and Galkin (EFG) \cite{Ebert:2011} (see also \cite{Chen:2014}). Indeed, the
quantum numbers $J^P=\frac52^+$ of $\Lambda_c(2880)$ have been correctly predicted in the model based on the diquark idea before the Belle experiment \cite{Selem}.
Moreover, EFG have shown that all available experimental data on heavy baryons fit nicely to the linear Regge trajectories, namely, the trajectories in the $(J,M^2)$ and $(n_r,M^2)$ planes for orbitally and radially excited heavy baryons, respectively:
\begin{eqnarray}
J=\alpha M^2+\alpha_0, \qquad n_r=\beta M^2+\beta_0,
\end{eqnarray}
where $n_r$ is the radial excitation quantum number, $\alpha$, $\beta$ are the slopes and $\alpha_0$, $\beta_0$ are intercepts. The Regge trajectories can be plotted for charmed baryons with natural $(P=(-1)^{J-1/2})$ and unnatural $(P=(-1)^{J+1/2})$ parties.
The linearity, parallelism and equidistance of the Regge trajectories were verified. The predictions of the spin-parity quantum numbers of charmed baryons and their masses in \cite{Ebert:2011} can be regarded as a theoretical benchmark. Specifically, the $J^P$ assignments are given by $\Lambda_c(2765): 1/2^+(2S); ~\Sigma_c(2800): 3/2^-(1P);~ \Xi'_c(2930): 1/2^-,3/2^-,5/2^-(1P);~ \Xi'_c(2980): 1/2^+(2S);~ \Xi_c(3055): 3/2^+(1D);~ \Xi'_c(3080): 5/2^+(1D);~ \Xi_c(3123): 7/2^+(1D)$.

Since the $J^P=1/2^-$ and $3/2^-$ antitriplets are well established (see Table \ref{tab:3and6}), one may wonder what are the counterparts in the ${\bf 6}$? It turns out that there is no $J^P={1\over 2}^-$ sextet as the $\Sigma_c(2800)$ cannot be assigned with such spin-parity quantum numbers. This should not be a surprise given that the light $\Sigma$ baryon with $J^P={1/2}^-$ also has not been seen \cite{PDG}.
The next possible sextet is for $J^P={3/2}^-$: ($\Omega_c(3050)^0$, $\Xi'_c(2930)^{+,0},\Sigma_c(2800)^{++,+,0}$) where the $\Omega_c({3/ 2}^-)$ is predicted to have a mass 3050 MeV by the quark-diquark model \cite{Ebert:2011}. The mass differences in this sextet, $\Delta m_{\Xi'_c\Sigma_c}=131$ MeV and $\Delta m_{\Omega_c\Xi'_c}=119$ MeV are consistent with that measured in $J^P=1/2^+$ and $3/2^+$ sextets (c.f. Table \ref{tab:3and6}).

On the basis of QCD sum rules, many charmed baryon multiplets classified according to $[{\bf 6}_F({\rm or~} {\bf \bar 3}_F),J_\ell,S_\ell, \rho/\lambda)]$ were studied in  \cite{Chen:2015k} with focus on the physics of $\rho$- and $\lambda$-mode excitations. Three sextets were proposed in this work: $(\Omega_c(3250),\Xi'_c(2980),\Sigma_c(2800))$ for $J^P=1/2^-,3/2^-$ and $(\Omega_c(3320),\Xi'_c(3080),\Sigma_c(2890))$ for $J^P=5/2^-$. Notice that
$\Xi'_c(2980)$ and $\Xi'_c(3080)$ were treated as $p$-wave baryons rather than first positive-parity excitations.
The results on the multiplet $[{\bf 6}_F,1,0,\rho]$ led the authors of \cite{Chen:2015k} to suggest that there are two $\Sigma_c(2800)$, $\Xi'_c(2980)$ and $\Omega_c(3250)$ states with $J^P=1/2^-$ and $J^P=3/2^-$. The mass splittings are $14\pm7$, $12\pm7$ and $10\pm6$ MeV, respectively. The predicted mass of $\Omega_c(1/2^-,3/2^-)$ around $3250\pm200$ MeV is to be compared with 3050 MeV calculated in the quark-diquark model. Using the central value of the predicted masses to label the states in the multiplet $[{\bf 6}_F,1,0,\rho]$ (see Table I of \cite{Chen:2015k}), one will have
\be
J^P={1/ 2}^-:&&(\Omega_c(3250),\Xi'_c(2960),\Sigma_c(2730)),~~\Delta m_{\Xi'_c\Sigma_c}=230\pm234,~\Delta m_{\Omega_c\Xi'_c}=290\pm250, \non \\
J^P={3/ 2}^-:&&(\Omega_c(3260),\Xi'_c(2980),\Sigma_c(2750)),~~\Delta m_{\Xi'_c\Sigma_c}=230\pm234,~\Delta m_{\Omega_c\Xi'_c}=280\pm242, \non \\
\en
in units of MeV.
Because of the large theoretical uncertainties in masses, it is not clear if the QCD sum-rule calculations are compatible with the mass differences measured in $J^P=1/2^+$ and $3/2^+$ sextets, namely, $\Delta m_{\Xi'_c\Sigma_c}\approx 125$ MeV and $\Delta m_{\Omega_c\Xi'_c}\approx 120$ MeV. At any rate, it will be interesting to test these two different model predictions for $J^P=3/2^-$ and $1/2^-$ sextets in the future.

Finally, we would like to remark that in recent years there have been intensive lattice studies of singly, doubly and triply charmed baryon spectra by many different groups; see e.g. \cite{Bali:2015,Padmanath:2015bra} and references therein. However, the current lattice QCD calculations on singly charmed baryons focus mostly on the low-lying $1/2^+$ and $3/2^+$ states. 
There exist some preliminary lattice results on excited charmed baryon spectroscopy, but the identification with observed charmed baryon states has not been made \cite{Padmanath:2015bra,Padmanath:2013bla}.
It will be very interesting if the lattice studies in the future can provide us information on the spin-parity quantum numbers of $p$-wave and $d$-wave excited states such as $\Lambda_c(2765),\Sigma_c(2800), \Xi_c(2980),\Xi_c(3055),\cdots$ and etc.

\section{Strong decays}

As stated in the Introduction, strong decays of charmed baryons involving soft pseudoscalar mesons are most conveniently described by HHChPT. The chiral Lagrangian involves two coupling constants
$g_1$ and $g_2$ for $P$-wave transitions between $s$-wave and
$s$-wave baryons \cite{Yan}, six couplings $h_{2}-h_7$ for the
$S$-wave transitions between $s$-wave and $p$-wave baryons, and
eight couplings $h_{8}-h_{15}$ for the $D$-wave transitions
between $s$-wave and $p$-wave baryons \cite{Pirjol}. The general
chiral Lagrangian for heavy baryons coupling to the pseudoscalar
mesons can be expressed compactly in terms of superfields. We will
not write down the relevant Lagrangians here; instead the reader
is referred to Eqs. (3.1) and (3.3) of \cite{Pirjol}.
Nevertheless, we list some of the partial widths derived from the
Lagrangian \cite{Pirjol}:

 \begin{eqnarray} \label{eq:swavecoupling}
 \Gamma(\Sigma_c^*\to \Sigma_c\pi)={g_1^2\over 2\pi
 f_\pi^2}\,{m_{\Sigma_c}\over m_{\Sigma_c^*}}p_\pi^3, &&
 \Gamma(\Sigma_c\to \Lambda_c\pi)={g_2^2\over 2\pi
 f_\pi^2}\,{m_{\Lambda_c}\over m_{\Sigma_c}}p_\pi^3, \nonumber \\
 \Gamma(\Lambda_{c1}(1/2^-)\to \Sigma_c\pi)={h_2^2\over 2\pi
 f_\pi^2}\,{m_{\Sigma_c}\over m_{\Lambda_{c1}}}E_\pi^2p_\pi, &&
 \Gamma(\Sigma_{c0}(1/2^-)\to \Lambda_c\pi)={h_3^2\over 2\pi
 f_\pi^2}\,{m_{\Lambda_c}\over m_{\Sigma_{c0}}}E_\pi^2p_\pi,
 \nonumber \\
 \Gamma(\Sigma_{c1}(1/2^-)\to \Sigma_c\pi)={h_4^2\over 4\pi
 f_\pi^2}\,{m_{\Sigma_c}\over m_{\Sigma_{c1}}}E_\pi^2p_\pi, &&
 \Gamma(\tilde\Sigma_{c1}(1/2^-)\to \Sigma_c\pi)={h_5^2\over 4\pi
 f_\pi^2}\,{m_{\Sigma_c}\over m_{\tilde\Sigma_{c1}}}E_\pi^2p_\pi,
 \nonumber \\
 \Gamma(\tilde\Xi_{c0}(1/2^-)\to \Xi_c\pi)={h_6^2\over 2\pi
 f_\pi^2}\,{m_{\Xi_c}\over m_{\tilde\Xi_{c0}}}E_\pi^2p_\pi, &&
 \Gamma(\tilde\Lambda_{c1}(1/2^-)\to \Sigma_c\pi)={h_7^2\over 2\pi
 f_\pi^2}\,{m_{\Sigma_c}\over m_{\tilde\Lambda_{c1}}}E_\pi^2p_\pi,
 \nonumber \\
  \Gamma(\Lambda_{c1}(3/2^-)\to\Sigma_c\pi)={2h_8^2\over 9\pi
 f_\pi^2}\,{m_{\Sigma_c}\over m_{\Lambda_{c1}(3/2)}}\,p_\pi^5, &&
  \Gamma\left(\Sigma_{c1}({3/2}^-)\to\Sigma_c^{(*)}\pi\right) =
 {h_{9}^2\over 9\pi f_\pi^2}\,{m_{\Sigma_c^{(*)}}\over
 m_{\Sigma_{c1}(3/2)}}p_\pi^5,
 \nonumber \\
 \Gamma\left(\Sigma_{c2}({3/ 2}^-)\to\Lambda_c\pi\right)
 = {4h_{10}^2\over 15\pi f_\pi^2}\,{m_{\Lambda_c}\over
 m_{\Sigma_{c2}}}p_\pi^5, &&
 \Gamma\left(\Sigma_{c2}({3/2}^-)\to\Sigma_c^{(*)}\pi\right) =
 {h_{11}^2\over 10\pi f_\pi^2}\,{m_{\Sigma_c^{(*)}}\over
 m_{\Sigma_{c2}}}p_\pi^5,
  \\
 \Gamma\left(\Sigma_{c2}({5/ 2}^-)\to\Sigma_c\pi\right)
 = {2h_{11}^2\over 45\pi f_\pi^2}\,{m_{\Sigma_c}\over
 m_{\Sigma_{c2}}}p_\pi^5, &&
 \Gamma\left(\Sigma_{c2}({5/2}^-)\to\Sigma_c^{*}\pi\right) =
 {7h_{11}^2\over 45\pi f_\pi^2}\,{m_{\Sigma_c^{*}}\over
 m_{\Sigma_{c2}}}p_\pi^5,
 \nonumber \\
 \Gamma\left(\tilde\Sigma_{c1}({3/ 2}^-)\to\Sigma_c\pi\right)
 = {h_{12}^2\over 9\pi f_\pi^2}\,{m_{\Sigma_c}\over
 m_{\tilde\Sigma_{c1}}}p_\pi^5, &&
 \Gamma\left(\tilde\Lambda_{c1}({3/2}^-)\to\Sigma_c\pi\right) =
 {4h_{13}^2\over 9\pi f_\pi^2}\,{m_{\Sigma_c}\over
 m_{\tilde\Lambda_{c1}}}p_\pi^5,
 \nonumber \\
 \Gamma\left(\tilde\Xi_{c2}({3/ 2}^-)\to\Xi_c\pi\right)
 = {4h_{14}^2\over 15\pi f_\pi^2}\,{m_{\Xi_c}\over
 m_{\tilde\Xi_{c2}}}p_\pi^5, &&
 \Gamma\left(\tilde\Lambda_{c2}({3/2}^-)\to\Sigma_c\pi\right) =
 {h_{15}^2\over 5\pi f_\pi^2}\,{m_{\Sigma_c}\over
 m_{\tilde\Lambda_{c2}}}p_\pi^5,
 \nonumber
 \end{eqnarray}
where $p_\pi$ is the pion's momentum and $f_\pi=132$ MeV. The dependence on the pion momentum is proportional to $p_\pi$, $p_\pi^3$ and $p_\pi^5$ for $S$-wave, $P$-wave and $D$-wave transitions, respectively. It is obvious that the couplings $g_1,g_2,h_2,\cdots,h_7$ are dimensionless, while $h_8,\cdots,h_{15}$ have canonical dimension $E^{-1}$.

\subsection{Strong decays of $s$-wave charmed baryons}
Since the strong decay $\Sigma_c^*\to\Sigma_c\pi$ is
kinematically prohibited,
the coupling $g_1$ cannot be extracted directly from the strong
decays of heavy baryons.
In the framework of HHChPT, one can use some measurements as input to fix the
coupling $g_2$ which, in turn, can be used to predict the rates of
other strong decays. Among the strong decays $\Sigma_c^{(*)}\to\Lambda_c\pi$, $\Sigma_c^{++}\to\Lambda_c^+\pi^+$ is the most well measured. Hence, we shall use this mode to extract the coupling $g_2$.

Using the 2006 data  of $\Gamma(\Sigma_c^{++})=\Gamma(\Sigma_c^{++}\to\Lambda_c^+\pi^+)=
2.23\pm0.30\,{\rm MeV}$ \cite{PDG2006}, we obtain the coupling $g_2$ to be
\begin{eqnarray} \label{eq:g2}
 |g_2|_{_{2006}}=0.605^{+0.039}_{-0.043}\,,
\end{eqnarray}
where we have neglected the tiny contributions from
electromagnetic decays.
The predicted rates of other modes are shown in Table \ref{tab:strongdecayS}, for example,
\begin{eqnarray}
\Gamma(\Xi_c^{'*+})=\Gamma(\Xi_c^{'*+}\to\Xi_c^+\pi^0,\Xi_c^0\pi^+)
&=& {g_2^2\over 4\pi
 f_\pi^2}\left({1\over 2}{m_{\Xi_c^+}\over m_{\Xi'^+_c}}p_\pi^3+
 {m_{\Xi_c^0}\over m_{\Xi'^+_c}}p_\pi^3\right)=
(2.8\pm 0.4)\,{\rm MeV},   \nonumber \\
\Gamma(\Xi_c^{'*0})=\Gamma(\Xi_c^{'*0}\to\Xi_c^+\pi^-,\Xi_c^0\pi^0)
&=& {g_2^2\over 4\pi
 f_\pi^2}\left({m_{\Xi_c^+}\over m_{\Xi'^0_c}}p_\pi^3+{1\over 2}
 {m_{\Xi_c^0}\over m_{\Xi'^0_c}}p_\pi^3\right)=
(2.9\pm 0.4)\,{\rm MeV}.
\end{eqnarray}
Note that we have neglected the effect of $\Xi_c-\Xi'_c$ mixing in
calculations (for recent considerations, see \cite{Boyd,Ito}).
It is clear from Table \ref{tab:strongdecayS} that the agreement between theory and experiment is excellent except the predicted width for $\Sigma_c^{*++}\to\Lambda_c^+\pi^+$ is a bit too large.

Using the new data from 2014 Particle Data Group \cite{PDG} in conjunction with the new measurements of $\Sigma_c$ and $\Sigma_c^*$ widths by Belle \cite{Belle:2014}, we have $\Gamma(\Sigma_c^{++}\to\Lambda_c^+\pi^+)=
1.94^{+0.08}_{-0.16}\,{\rm MeV}$. Therefore, the coupling $g_2$ is reduced to
\begin{eqnarray} \label{eq:newg2}
 |g_2|_{_{2015}}=0.565^{+0.011}_{-0.024}\,.
\end{eqnarray}
From Table \ref{tab:strongdecayS} we see that the agreement between theory and experiment is further improved: The predicted $\Xi_c(2645)^+$ width is consistent with the first new measurement by Belle \cite{Belle:dc} and the new calculated width for $\Sigma_c^{*++}\to\Lambda_c^+\pi^+$ is now in agreement with experiment.
It is also clear that the $\Sigma_c$ width is
smaller than that of $\Sigma_c^*$ by a factor of $\sim 7$,
although they will become the same in the limit of heavy quark
symmetry. This is ascribed to the fact that the pion's momentum is
around 90 MeV in the decay $\Sigma_c\to\Lambda_c\pi$ while it is
two times bigger in $\Sigma_c^*\to\Lambda_c\pi$.

\begin{table}[t]
\caption{Decay widths (in units of MeV) of $s$-wave charmed
baryons.  Data under the label Expt.(2015) are taken from 2014 PDG \cite{PDG} together with the new measurements of $\Sigma_c$, $\Sigma_c^*$ \cite{Belle:2014}
and $\Xi_c(2645)^+$ widths \cite{Belle:dc}.} \label{tab:strongdecayS}
\begin{center}
\begin{tabular}{|c|c c||c c|} \hline \hline
~~~~~~~Decay~~~~~~~ & ~Expt.(2006)~ & ~HHChPT(2006)~ & ~Expt.(2015)~ & ~HHChPT(2015)~  \\
\hline
 $\Sigma_c^{++}\to\Lambda_c^+\pi^+$ & $2.23\pm0.30$ & input & $1.94^{+0.08}_{-0.16}$ & input  \\ \hline
 $\Sigma_c^{+}\to\Lambda_c^+\pi^0$ & $<4.6$ & $2.6\pm0.4$ & $<4.6$ & $2.3^{+0.1}_{-0.2}$ \\ \hline
 $\Sigma_c^{0}\to\Lambda_c^+\pi^-$ & $2.2\pm0.4$ & $2.2\pm0.3$ & $1.9^{+0.1}_{-0.2}$ & $1.9^{+0.1}_{-0.2}$ \\ \hline
 $\Sigma_c(2520)^{++}\to\Lambda_c^+\pi^+$ & $14.9\pm1.9$ & $16.7\pm2.3$ & $14.8^{+0.3}_{-0.4}$ & $14.5^{+0.5}_{-0.8}$  \\  \hline
 $\Sigma_c(2520)^{+}\to\Lambda_c^+\pi^0$ & $<17$ & $17.4\pm2.3$  & $<17$ & $15.2^{+0.6}_{-1.3}$ \\  \hline
 $\Sigma_c(2520)^{0}\to\Lambda_c^+\pi^-$ & $16.1\pm2.1$ & $16.6\pm2.2$  & $15.3^{+0.4}_{-0.5}$ & $14.7^{+0.6}_{-1.2}$  \\  \hline
 $\Xi_c(2645)^+\to\Xi_c^{0,+}\pi^{+,0}$ & $<3.1$ & $2.8\pm0.4$ & $2.6\pm0.5$ & $2.4^{+0.1}_{-0.2}$   \\  \hline
 $\Xi_c(2645)^0\to\Xi_c^{+,0}\pi^{-,0}$ & $<5.5$ & $2.9\pm0.4$ & $<5.5$ & $2.5^{+0.1}_{-0.2}$  \\ \hline \hline
\end{tabular}
\end{center}
\end{table}

It is worth remarking that although the coupling $g_1$ cannot be
determined directly from the strong decay such as
$\Sigma_c^*\to\Sigma_c\pi$, some information of $g_1$ can be
learned from the radiative decay $\Xi_c^{'*0}\to\Xi_c^0 \gamma$,
which is prohibited at tree level by SU(3) symmetry but can be
induced by chiral loops. A measurement of
$\Gamma(\Xi_c^{'*0}\to\Xi_c^0\gamma)$ will yield two possible
solutions for $g_1$.
As pointed out in \cite{Yan}, within the framework of the
non-relativistic quark model, the couplings $g_1$ and $g_2$ can be
related to $g_A^q$, the axial-vector coupling in a single quark
transition of $u\to d$, via
\begin{eqnarray}
 g_1={4\over 3}g_A^q, \qquad\qquad g_2=\sqrt{2\over 3}g_A^q.
\end{eqnarray}
Assuming the validity of the quark model
relations among different coupling constants, the experimental
value of $g_2$ implies $|g_1|=0.93\pm 0.16$ \cite{Cheng97}.

The couplings $g_1$ and $g_2$ have been evaluated using lattice QCD with the results \cite{Detmold} \footnote{Our definitions of $g_1$ and $g_2$ are related to that of Detmold, Lin and Meinel \cite{Detmold} by the relations: $g_1=(2/3)g_2^{\rm DLM}$ and $g_2=g_3^{\rm DLM}/\sqrt{3}$.}
\begin{eqnarray}
g_1=0.56\pm0.13\,,\qquad\quad g_2=0.41\pm0.08\,.
\end{eqnarray}
Hence, the quark model values of $g_1$ and $g_2$ are significantly larger than the above lattice QCD results. This is ascribed to the fact that $1/m_Q$ corrections to strong decays have been taken into account in lattice calculations \cite{Detmold}. For example, $1/m_c$ effect on the amplitude of $\Sigma_c^{(*)}\to \Lambda_c\pi$ is about 40\%. As a consequence, the lattice values of of $g_1$ and $g_2$ are significantly smaller than the quark model results.

\subsection{Strong decays of $p$-wave charmed baryons}

As noted in passing, six couplings $h_{2}-h_7$ are needed to describe the
$S$-wave transitions between $s$-wave and $p$-wave baryons, and
eight couplings $h_{8}-h_{15}$ for the $D$-wave transitions
between $s$-wave and $p$-wave baryons \cite{Pirjol}.
Since $\Lambda_c(2595)^+$ and $\Lambda_c(2625)^+$ form a doublet
$\Lambda_{c1}({1\over 2}^-,{3\over 2}^-)$, the couplings $h_2$ and $h_8$ in principle can be extracted from $\Lambda_c(2595)\to
\Sigma_c\pi$ and $\Lambda_c(2625)\to
\Sigma_c\pi$, respectively. However, this method is not practical as only the decay $\Lambda_c(2595)^+\to\Sigma^+\pi^0$ is kinematically (barely) allowed (see discussions below), while the $\Lambda_c(2625)$ decay to $\Sigma_c\pi$ via a $D$-wave transition is kinematically suppressed.

Likewise, the information on the couplings $h_{10}$ and $h_{11}$  can be inferred from the strong decays of $\Sigma_c(2800)$ identified with $\Sigma_{c2}(3/2^-)$.
Couplings other than $h_2$, $h_8$ and $h_{10}$ can be related to each
other via the quark model. The $S$-wave couplings between the
$s$-wave and the $p$-wave baryons are related by \cite{Pirjol}
 \begin{eqnarray} \label{eq:QMh3}
 {|h_3|\over |h_4|}={\sqrt{3}\over 2}, \quad {|h_2|\over |h_4|}={1\over 2},
 \quad {|h_5|\over |h_6|}={2\over \sqrt{3}},\quad {|h_5|\over
 |h_7|}=1\,.
 \end{eqnarray}
The $D$-wave couplings satisfy the relations
 \begin{eqnarray} \label{eq:QMh8}
 |h_8|=|h_9|=|h_{10}|, \quad {|h_{11}|\over |h_{10}|}={|h_{15}|\over |h_{14}|}=\sqrt{2},
 \quad {|h_{12}|\over |h_{13}|}=2, \quad {|h_{14}|\over
 |h_{13}|}=1\,.
 \end{eqnarray}
The reader is referred to \cite{Pirjol} for further details.

Although the coupling $h_2$ can be inferred from the two-body decay $\Lambda_c(2595)\to\Sigma_c\pi$, this method is less accurate because the decay is kinematically barely allowed or even prohibited depending on the mass of $\Lambda_c(2595)^+$.
Since
$m(\Sigma_c^{++})+m(\pi^-)=2593.55$ MeV, $m(\Sigma_c^{+})+m(\pi^0)=2587.88$ MeV and  $m(\Sigma_c^{0})+m(\pi^+)=2593.31$ MeV, it is obvious that the decays $\Lambda_c(2595)^+\to\Sigma_c^{++}\pi^-,\Sigma_c^0\pi^+$ and $\Lambda_c(2595)^+\to\Sigma^+\pi^0$ are kinematically barely allowed for $m(\Lambda_c(2595))=2595.4$ MeV, while only the last mode is allowed
for $m(\Lambda_c(2595))=2592.25$ MeV. Moreover, the finite width effect of the intermediate resonant states could become important \cite{Falk03}.

We next turn to the three-body decays $\Lambda_c^+\pi\pi$ of  $\Lambda_c(2595)^+$ and $\Lambda_c(2625)^+$ to extract $h_2$ and $h_8$. Since the 3-body decay of the latter proceeds in a $P$-wave, it is expected to be suppressed.
Using the measured ratios of $\Gamma(\Lambda_c(2595)^+\to\Sigma_c^{++}\pi^-)$ and $\Gamma(\Lambda_c(2595)^+\to\Sigma_c^{0}\pi^+)$ relative to $\Gamma(\Lambda_c(2595)^+\to\Lambda_c^+\pi^+\pi^-)$, assuming $\Gamma(\Lambda_c(2595)^+\to\Lambda_c^+\pi^0\pi^0)\approx\Gamma(\Lambda_c(2595)^+\to\Lambda_c^+\pi^+\pi^-)$ (see the discussion below for the justification) and using the 2006 data from PDG \cite{PDG2006} for $\Gamma(\Lambda_c(2595))=3.6^{+2.0}_{-1.3}$ MeV and the mass $m(\Lambda_c(2595))=2595.4\pm0.6$ MeV, we have obtained the experimental resonant rate \cite{CC}
\be
\Gamma(\Lambda_c(2593)^+\to\Lambda_c^+\pi\pi)_R &=& (2.63^{+1.56}_{-1.09})\,{\rm
 MeV}
\en
as shown in Table \ref{tab:strongdecayP}.

Assuming the pole contributions to $\Lambda_c(2595)^+\to \Lambda_c^+\pi\pi$ due to the intermediate states $\Sigma_c$ and $\Sigma_c^*$, the resonant rate
for the process
$\Lambda_{c_1}^+(2595)\to \Lambda_c^+\pi^+\pi^-$ can be calculated
in the framework of HHChPT to be
\cite{Pirjol}
 \be
& & {d^2\Gamma(\Lambda_{c1}^{+}(2595)\to
\Lambda_c^+\pi^+(E_1)\pi^-(E_2))\over dE_1dE_2}=  \\
& &\qquad \frac{g_2^2}{16\pi^3 f_\pi^4}m_{\Lambda_c^+}\left\{ {\bf
p}_2^2|A|^2 + {\bf p}_1^2|B|^2 + 2{\bf p}_1\cdot{\bf p}_2\,
\mbox{Re }(AB^*)\right\},  \nonumber
 \en
with
 \be
 && A(E_1,E_2)
 = \frac{h_2E_1}{\Delta_R-\Delta_{\Sigma_c^0}-E_1+
i\Gamma_{\Sigma_c^0}/2}-\frac{\frac23 h_8{\bf p}_2^2}
{\Delta_R-\Delta_{\Sigma_c^{*0}}-E_1+i\Gamma_{\Sigma_c^{*0}}/2}, \\
& & \qquad\qquad  +\frac{2h_8{\bf p}_1\cdot{\bf p}_2}
{\Delta_R-\Delta_{\Sigma_c^{*++}}-E_2+i\Gamma_{\Sigma_c^{*++}}/2}\,,
\nonumber\\
&& B(E_1,E_2;\Delta_{\Sigma_c^{(*)0}},\Delta_{\Sigma_c^{(*)++}}) =
A(E_2,E_1;\Delta_{\Sigma_c^{(*)++}},\Delta_{\Sigma_c^{(*)0}})\,,
 \en
where $\Delta _R=m_{\Lambda_c(2593)}-m_{\Lambda_c}$ and $\Delta
_{\Sigma_c^{(*)}}=m_{\Sigma_c^{(*)}}-m_{\Lambda_c}$. Likewise, a similar relation can be derived for
$\Lambda_{c_1}^+(2625)\to \Lambda_c^+\pi^+\pi^-$ (see \cite{Pirjol}). Numerically, we found \cite{CC}
\begin{eqnarray} \label{eq:h2h8}
 \Gamma(\Lambda_c(2595)^+\to\Lambda_c^+\pi\pi)_R&=& 13.82h_2^2
 +26.28h_8^2-2.97h_2h_8, \nonumber \\
 \Gamma(\Lambda_c(2625)^+\to\Lambda_c^+\pi\pi)_R&=& 0.617h_2^2+0.136\times
 10^6h_8^2-27h_2h_8,
\end{eqnarray}
where $\Lambda_c^+\pi\pi=\Lambda_c^+\pi^+\pi^- +\Lambda_c^+\pi^0\pi^0$.
It is clear that the limit on $\Gamma(\Lambda_c(2625))$ gives an
upper bound on $h_8$ of order $10^{-3}$ (in units of MeV$^{-1}$),
whereas the decay width of $\Lambda_c(2595)$ is entirely governed
by the coupling $h_2$
\begin{eqnarray} \label{eq:h2fw}
 |h_2|_{_{\rm 2006}}=0.437^{+0.114}_{-0.102}\,, \qquad\quad |h_8|_{_{\rm 2006}}< 3.65\times
 10^{-3}\,{\rm MeV}^{-1}\,.
\end{eqnarray}

It was pointed out in \cite{Falk03} that the proximity of the $\Lambda_c(2595)^+$ mass to the sum of the masses of its decay products will lead to an important threshold effect which will lower the $\Lambda_c(2595)^+$ mass by $2-3$ MeV than the one observed. A more sophisticated treatment of the mass lineshape of $\Lambda_c(2595)^+\to\Lambda_c^+\pi^+\pi^-$ by CDF  with data sample 25 times larger than previous measurements yields $m(\Lambda_c(2595))=2592.25\pm0.28$ MeV \cite{CDF:2595}, which is 3.1 MeV smaller than the 2006 world average.
Therefore, strong decays of $\Lambda_c(2595)$ into $\Lambda_c\pi\pi$ are very close to the threshold as $m_{\Lambda_c(2595)}-m_{\Lambda_c}=305.79\pm0.24$ MeV \cite{PDG}.
Hence, its phase space is very sensitive to the small
isospin-violating mass differences between members of pions and
charmed Sigma baryon multiplets.

\begin{table}[t]
\caption{Same as Table \ref{tab:strongdecayS} except for $p$-wave
charmed baryons. The results under the label HHChPT(2015) are obtained using $g_2=0.565$, $h_2=0.63\pm0.07$ and $h_8= (0.85^{+0.11}_{-0.08})\times 10^{-3}{\rm MeV}^{-1}$. } \label{tab:strongdecayP}
\begin{center}
\begin{tabular}{|c|c c|| c c| } \hline \hline
~~~~~~~Decay~~~~~~~ & ~Expt.(2006)~ & ~HHChPT(2006)~ & ~Expt.(2015)~ & ~HHChPT(2015)~  \\ \hline
 $\Lambda_c(2595)^+\to (\Lambda_c^{+}\pi\pi)_R$ & $2.63^{+1.56}_{-1.09}$ & input  & $2.59\pm0.56$ & input \\ \hline
 $\Lambda_c(2595)^+\to \Sigma_c^{++}\pi^-$ & $0.65^{+0.41}_{-0.31}$ & $0.72^{+0.43}_{-0.30}$  & & \\ \hline
 $\Lambda_c(2595)^+\to \Sigma_c^{0}\pi^+$ & $0.67^{+0.41}_{-0.31}$ & $0.77^{+0.46}_{-0.32}$ & & \\ \hline
 $\Lambda_c(2595)^+\to \Sigma_c^{+}\pi^0$ & & $1.57^{+0.93}_{-0.65}$ & & $2.38^{+0.56}_{-0.50}$   \\ \hline
 $\Lambda_c(2625)^+\to \Sigma_c^{++}\pi^-$ & $<0.58$ & $0.029$ & $<0.30$ & $ 0.028$ \\ \hline
 $\Lambda_c(2625)^+\to \Sigma_c^{0}\pi^+$ & $<0.60$ & $0.029$ & $<0.30$ & $0.029$  \\ \hline
 $\Lambda_c(2625)^+\to \Sigma_c^{+}\pi^0$ & & $ 0.041$ & & $ 0.040$  \\ \hline
 $\Lambda_c(2625)^+\to \Lambda_c^+\pi\pi$ & $<1.9$ & $ 0.21$ & $<0.97$ & $ 0.32$ \\ \hline
 $\Sigma_c(2800)^{++}\to\Lambda_c\pi,\Sigma_c^{(*)}\pi$ & $75^{+22}_{-17}$ & input  & $75^{+22}_{-17}$ & input \\ \hline
 $\Sigma_c(2800)^{+}\to\Lambda_c\pi,\Sigma_c^{(*)}\pi$ & $62^{+60}_{-40}$ & input  & $62^{+60}_{-40}$ & input \\ \hline
 $\Sigma_c(2800)^0\to\Lambda_c\pi,\Sigma_c^{(*)}\pi$ & $61^{+28}_{-18}$ & input & $72^{+22}_{-15}$ & input  \\ \hline
 $\Xi_c(2790)^+\to\Xi'^{0,+}_c\pi^{+,0}$ & $<15$ & $8.0^{+4.7}_{-3.3}$ & $<15$ & $16.7^{+3.6}_{-3.6}$ \\ \hline
 $\Xi_c(2790)^0\to\Xi'^{+,0}_c\pi^{-,0}$  & $<12$ & $8.5^{+5.0}_{-3.5}$ & $<12$ & $17.7^{+2.9}_{-3.8}$ \\ \hline
 $\Xi_c(2815)^+\to\Xi^{*+,0}_c\pi^{0,+}$ & $<3.5$ & $3.4^{+2.0}_{-1.4}$ & $<3.5$ & $7.1^{+1.5}_{-1.5}$ \\ \hline
 $\Xi_c(2815)^0\to\Xi^{*+,0}_c\pi^{-,0}$ & $<6.5$ & $3.6^{+2.1}_{-1.5}$ & $<6.5$ &  $7.7^{+1.7}_{-1.7}$ \\ \hline \hline
\end{tabular}
\end{center}
\end{table}

For $m(\Lambda_c(2595))=2592.25\pm0.28$ MeV \cite{CDF:2595} we obtain (in units of MeV)
\begin{eqnarray} \label{eq:h2h8new}
 \Gamma(\Lambda_c(2595)^+\to\Lambda_c^+\pi\pi)_R &=& g_2^2(20.45h_2^2
 +43.92h_8^2-8.95h_2h_8), \nonumber \\
 \Gamma(\Lambda_c(2625)^+\to\Lambda_c^+\pi\pi)_R &=& g_2^2(1.78h_2^2+4.557\times
 10^6h_8^2-79.75h_2h_8).
\end{eqnarray}
By performing a fit to the measured $M(pK^-\pi^+\pi^+)-M(pK^-\pi^+)$ and $M(pK^-\pi^+\pi^+\pi^-)-M(pK^-\pi^+)$ mass difference distributions and using $g_2^2=0.365$, CDF found $h_2^2=0.36\pm0.08$ or $|h_2|=0.60\pm0.07$ \cite{CDF:2595}. This corresponds to a decay width $\Gamma(\Lambda_c(2595)^+)=2.59\pm0.30\pm0.47$ MeV \cite{CDF:2595}. \footnote{Note that the contributions from the $h_8$ terms to $\Gamma(\{\Lambda_c(2595)^+,\Lambda_c(2625)^+\}\to\Lambda_c^+\pi\pi)$ have been ignored in the CDF fit to the data \cite{CDF:2595}. Using $g_2^2=0.365$ and $h_2^2=0.36\pm0.08$, we obtain $\Gamma(\Lambda_c(2595)^+)\approx 2.68$ MeV from Eq. (\ref{eq:h2h8new}), which is slightly larger than the CDF value of $2.59\pm0.56$ MeV. This is mainly because we have used the updated widths for $\Sigma_c$ and $\Sigma_c^*$.}
Note that the decay width of $\Lambda_c(2595)^+$ measured by CDF is the quantity $\Gamma(\Lambda_c(2595)^+\to\Lambda_c^+\pi\pi)_R$ instead of the natural width associated with a Breit-Wigner curve. For the width of $\Lambda_c(2625)^+$, CDF observed a value consistent with zero and therefore calculated an upper limit 0.97 MeV using a Bayesian approach. According to PDG, $\Lambda_c(2625)^+\to\Lambda_c^+\pi\pi$
is dominated by direct nonresonant contributions \cite{PDG}.

From the CDF measurements $\Gamma(\Lambda_c(2595)^+)=2.59\pm0.56$ MeV and $\Gamma(\Lambda_c(2625)^+)<0.97$ MeV, we obtain
\begin{eqnarray} \label{eq:h2h8,2014}
 |h_2|_{_{\rm 2015}}=0.63\pm0.07\,, \qquad\quad |h_8|_{_{\rm 2015}}< 2.32\times  10^{-3}\,{\rm MeV}^{-1}\,.
\end{eqnarray}
Hence, the magnitude of the coupling $h_2$ is greatly enhanced from 0.437 to 0.63\,. Our $h_2$ is slightly different from the value of 0.60 obtained by CDF. This is because CDF used $|g_2|=0.604$ to calculate the mass dependence of $\Gamma(\Lambda_c^+\pi\pi)$, while we used $|g_2|=0.565$. Since $\Gamma(\Lambda_c(2595)^+\to\Lambda_c^+\pi\pi)$ is basically proportional to $g_2^2h_2^2$, a smaller $g_2$ will lead to a larger $h_2$.

\begin{figure}[t]
\centerline{\psfig{file=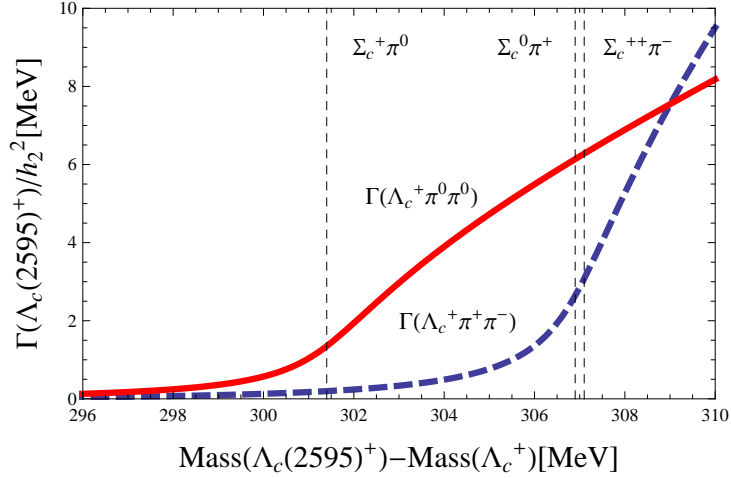,width=3.8in}}
\caption{Calculated dependence of $\Gamma(\Lambda_c^+\pi^0\pi^0)/h^2_2$ (full curve) and $\Gamma(\Lambda_c^+\pi^+\pi^-)/h^2_2$ (dashed curve) on $m(\Lambda_c(2595)^+)-m(\Lambda_c^+)$, where we have used the parameters $g_2=0.565$, $h_2=0.63$ and $h_8=0.85\times 10^{-3}\,{\rm MeV}^{-1}$. } \label{fig:Lambdac2595}
\end{figure}

The reader may wonder why the coupling $h_2$ obtained in 2006 and 2015 is so different even though the resonant rate of $\Lambda_c(2595)^+\to \Lambda_c^+\pi\pi$ used in 2006 and 2015 is very similar in its central value. This is ascribed to the fact that the mass of $\Lambda_c(2595)^+$ is 3.1 MeV lower than the previous world average due to the threshold effect. To illustrate this, following \cite{CDF:2595} we consider the dependence of $\Gamma(\Lambda_c^+\pi^+\pi^-)/h^2_2$ and $\Gamma(\Lambda_c^+\pi^0\pi^0)/h^2_2$ on $\Delta M(\Lambda_c(2595))\equiv M(\Lambda_c(2595)^+)-M(\Lambda_c^+)$ as depicted in Fig. \ref{fig:Lambdac2595}. For $\Delta M(\Lambda_c(2595))=308.9$ MeV, we see that $\Gamma(\Lambda_c^+\pi^0\pi^0)\approx \Gamma(\Lambda_c^+\pi^+\pi^-)$, while  $\Gamma(\Lambda_c^+\pi^0\pi^0)\approx 4.5\,\Gamma(\Lambda_c^+\pi^+\pi^-)$ for $\Delta M(\Lambda_c(2595))=305.79$ MeV. Due to the threshold effect, the isospin relation $\Gamma(\Lambda_c^+\pi^0\pi^0)={1\over 2} \Gamma(\Lambda_c^+\pi^+\pi^-)$ as advocated by PDG is strongly violated in $\Lambda_c(2595)\to\Lambda_c\pi\pi$ decays though it is still valid in $\Lambda_c(2625)\to\Lambda_c\pi\pi$ decays.
It is evident from Fig. \ref{fig:Lambdac2595} that $\Gamma(\Lambda_c^+\pi\pi)/h^2_2$ at $\Delta M(\Lambda_c(2595))=305.79$ MeV is smaller than that at 308.9 MeV. This explains why $h_2$ should become larger when $\Delta M(\Lambda_c(2595))$ becomes smaller.

The $\Xi_c(2790)$ and $\Xi_c(2815)$ baryons form a doublet
$\Xi_{c1}({1\over 2}^-,{3\over 2}^-)$. $\Xi_c(2790)$ decays to
$\Xi'_c\pi$, while $\Xi_c(2815)$ decays to $\Xi_c\pi\pi$,
resonating through $\Xi^*_c$, i.e. $\Xi_c(2645)$. Using the
coupling $h_2$ obtained from (\ref{eq:h2h8,2014}) and  assuming SU(3) flavor symmetry, the predicted $\Xi_c(2790)$ and $\Xi_c(2815)$
widths are shown in Table \ref{tab:strongdecayP}, where uses have
been made of
 \begin{eqnarray}
\Gamma(\Xi_{c1}(1/2^-)^+)\approx\Gamma(\Xi_{c1}(1/2^-)^+\to\Xi'^+_c\pi^0,\Xi'^0_c\pi^+)
&=& {h_2^2\over 4\pi
 f_\pi^2}\left({1\over 2}{m_{\Xi'^+_c}\over m_{\Xi_{c1}(1/2)}}E_\pi^2p_\pi+
 {m_{\Xi'^0_c}\over m_{\Xi_{c1}(1/2)}}E_\pi^2p_\pi\right),   \nonumber \\
\Gamma(\Xi_{c1}(3/2^-)^+)\approx\Gamma(\Xi_{c1}(3/2^-)^+\to\Xi^{*+}_c\pi^0,\Xi^{*0}_c\pi^+)
&=& {h_2^2\over 4\pi
 f_\pi^2}\left(\frac12{m_{\Xi^{*+}_c}\over m_{\Xi_{c1}(3/2)}}E_\pi^2p_\pi+
 {m_{\Xi^{*0}_c}\over m_{\Xi_{c1}(3/2)}}E_\pi^2p_\pi\right), \nonumber \\
 \end{eqnarray}
and similar expressions for the neutral
$\Xi_{c1}(\frac12^-,\frac32^-)$ states based on the experimental
observation that the $\Xi_c\pi\pi$ mode in $\Xi_c(2815)$ decays is
consistent with being entirely via $\Xi^*_c\pi$
\cite{CLEO:Xic2815}. It is evident that the predicted two-body decay rates of $\Xi_c(2790)^0$ and $\Xi_c(2815)^+$ exceed the current experimental limits because of the enhancement of $h_2$ (see Table \ref{tab:strongdecayP}). Hence, there is a tension for the coupling $h_2$ as its value extracted from from $\Lambda_c(2595)^+\to \Lambda_c^+\pi\pi$ will imply $\Xi_c(2790)^0\to\Xi'_c\pi$ and $\Xi_c(2815)^+\to\Xi_c^*\pi$ rates slightly above current limits. 
It is conceivable that SU(3) flavor symmetry breaking can help account for the discrepancy. For example, if $h_2$ is allowed to have $25\%$ uncertainties due to SU(3) breaking between $\Xi_{c1}$ and $\Lambda_{c1}$ decays, one will have $\Gamma(\Xi_c(2790)^0)\approx 9.9$ MeV and $\Gamma(\Xi_c(2815)^+)\approx 4.0$ MeV for $h_2=0.47$. It is clear that the former is consistent with the measured limit, while the discrepancy between theory and experiment for the latter is much improved. In HHChPT, SU(3) breaking effects arise from chiral loops due to the light quark masses. Applications to the strong decays of heavy baryons have been considered in \cite{Cheng:1993kp}. We plan to pursue this issue in the future.

Some information on the coupling $h_{10}$ can be inferred from
the strong decays of $\Sigma_c(2800)$. As noticed in passing, the
states $\Sigma_c(2800)^{++,+,0}$ are most likely to be
$\Sigma_{c2}({3\over 2}^-)$. From Eq. (\ref{eq:swavecoupling}) and the quark model relation $|h_3|=\sqrt{3}|h_2|$ from Eq. (\ref{eq:QMh3}), we obtain, for example, $\Gamma(\Sigma_{c0}^{++}\to\Lambda_c^+\pi^+)\approx 885$ MeV. Hence, $\Sigma_c(2800)$ cannot be identified with $\Sigma_{c0}(1/2^-)$.
Assuming their widths are dominated by the two-body
$D$-wave modes $\Lambda_c\pi$, $\Sigma_c\pi$ and $\Sigma_c^*\pi$,
we have \cite{Pirjol}
 \begin{eqnarray}
\Gamma\left(\Sigma_{c2}({3/2})^{++}\right) &\approx&
\Gamma\left(\Sigma_{c2}({3/ 2})^{++}\to\Lambda_c^+\pi^+\right)
\nonumber \\ &+&
\Gamma\left(\Sigma_{c2}({3/2})^{++}\to\Sigma_c^+\pi^+\right)+
\Gamma\left(\Sigma_{c2}({3/ 2})^{++}\to\Sigma_c^{*+}\pi^+\right)
\nonumber \\ &+&
\Gamma\left(\Sigma_{c2}({3/2})^{++}\to\Sigma_c^{++}\pi^0\right)+
\Gamma\left(\Sigma_{c2}({3/ 2})^{++}\to\Sigma_c^{*++}\pi^0\right),
 \end{eqnarray}
and similar expressions for $\Sigma_{c2}(\frac32)^+$ and
$\Sigma_{c2}(\frac32)^0$. The first term is governed by the $h_{10}^2$ coupling and the rest by $h_{11}^2$.
Using Eq. (\ref{eq:swavecoupling}), the quark model relation $h_{11}^2=2h_{10}^2$ [cf. Eq.
(\ref{eq:QMh8})] and the measured widths of
$\Sigma_c(2800)^{++,+,0}$ (Table \ref{tab:spectrum}), we obtain
 \begin{eqnarray}
|h_{10}|=(0.85^{+0.11}_{-0.08})\times 10^{-3}\,{\rm MeV}^{-1}\,.
 \end{eqnarray}
Roughly speaking, the $\Sigma_c(2800)$ widths are about 55 MeV and 15 MeV due to $\Lambda_c\pi$ and $\Sigma_c^{(*)}\pi$, respectively. Hence, the strong decays of $\Sigma_c(2800)$ are indeed dominated by the $\Lambda_c\pi$ mode.

The quark model relation $|h_8|=|h_{10}|$ then leads to \begin{eqnarray}
|h_{8}|\approx (0.85^{+0.11}_{-0.08})\times 10^{-3}\,{\rm MeV}^{-1}\,,
\end{eqnarray}
which improves the previous limit (\ref{eq:h2h8,2014}) by a factor of
3. The calculated partial widths of $\Lambda_c(2625)^+$
shown in Table \ref{tab:strongdecayP} are consistent with experimental limits.

\section{Conclusions}

We began with a brief overview of the charmed baryon spectroscopy and discussed their possible structure and $J^P$ assignment in the quark model.
We have assigned $\Sigma_{c2}({3\over 2}^-)$ to $\Sigma_c(2800)$. As for first positive-parity excitations,
with the help of the relativistic quark-diquark model and the $^3P_0$ model, we have identified $\tilde\Lambda_{c3}^2({5\over 2}^+)$ with $\Lambda_c(2800)$, $\tilde\Xi_{c}'({1\over 2}^+)$ with $\Xi_c(2980)$, and $\tilde\Xi_{c3}^2({5\over 2}^+)$ with $\Xi_c(3080)$, though the last assignment may encounter some potential problems.

With the new Belle measurement of the $\Sigma_c(2455)$ and $\Sigma_c(2520)$ widths and the recent CDF measurement of the strong decays of $\Lambda_c(2595)$ and $\Lambda_c(2625)$, we have updated coupling constants in heavy hadron chiral perturbation theory. We found $g_2=0.565^{+0.011}_{-0.024}$ for $P$-wave transition between $s$-wave and $s$-wave baryons, and $h_2=0.63\pm0.07$ extracted from $\Lambda_c(2595)^+\to\Lambda_c^+\pi\pi$. It is substantially enhanced compare to the old value of order 0.437. With the help from the quark model, two of the couplings $h_{10}$ and $h_{11}$ responsible for $D$-wave transitions between $s$-wave and $p$-wave baryons are determined from $\Sigma_c(2880)$ decays.  There is a tension for the coupling $h_2$ as its value extracted from $\Lambda_c(2595)^+\to \Lambda_c^+\pi\pi$ will $\Xi_c(2790)^0\to\Xi'_c\pi$ and $\Xi_c(2815)^+\to\Xi_c^*\pi$ rates slightly above the current limits. It is conceivable that SU(3) flavor symmetry breaking can help account for the discrepancy.

\section{Acknowledgments}
This research was supported in part by the Ministry of Science and Technology of R.O.C. under Grant
Nos. 104-2112-M-001-022 and  103-2112-M-033-002-MY3.

\newcommand{\bi}{\bibitem}

\end{document}